\documentclass[12pt]{article}

\usepackage[colorlinks,linkcolor=blue,citecolor=blue,urlcolor=blue,bookmarks,bookmarksnumbered]{hyperref}
\usepackage{amsmath,amssymb,amsfonts}
\usepackage{slashed,extarrows,bigints,cancel}
\usepackage[paper=letterpaper,margin=1.0in]{geometry}
\usepackage{graphicx,xcolor}

\usepackage{cite}
\newcommand{\be}{\begin{equation}}
\newcommand{\ee}{\end{equation}}
\newcommand{\bea}{\begin{eqnarray}}
\newcommand{\eea}{\end{eqnarray}}
\newcommand{\ba}{\begin{array}}
\newcommand{\ea}{\end{array}}
\newcommand{\ben}{\begin{enumerate}}
\newcommand{\een}{\end{enumerate}}
\newcommand{\bi}{\begin{itemize}}
\newcommand{\ei}{\end{itemize}}
\newcommand{\bc}{\begin{center}}
\newcommand{\ec}{\end{center}}
\newcommand{\bfig}{\begin{figure}}
\newcommand{\efig}{\end{figure}}
\newcommand{\bq}{\begin{quotation}}
\newcommand{\eq}{\end{quotation}}
\newcommand{\bt}{\begin{table}}
\newcommand{\et}{\end{table}}
\newcommand{\btab}{\begin{tabular}}
\newcommand{\etab}{\end{tabular}}
\newcommand{\bs}{\begin{slide}}
\newcommand{\es}{\end{slide}}

\newcommand{\IR}{\mathbb{R}}

\newcommand{\beq}{\begin{eqnarray}}
\newcommand{\eeq}{\end{eqnarray}}
\newcommand{\beqn}{\begin{eqnarray}}
\newcommand{\eeqn}{\end{eqnarray}}

\let\ba=\overline

\def\IR{\relax\leavevmode{\rm I\kern-.18em R}}
\def\ZZ{\relax\leavevmode
       \ifmmode\mathchoice
       {\hbox{\sf Z\kern-.4em Z}}
       {\hbox{\sf Z\kern-.4em Z}}
       {\lower.9pt\hbox{\scriptsize\sf Z\kern-.36em Z}}
       {\lower1.2pt\hbox{\tiny\sf Z\kern-.36em Z}}
       \else{\sf Z\kern-.4em Z}\fi}

\def\resetby#1#2{\@addtoreset{#2}{#1}}
\def\seceq{\@addtoreset{equation}{section}
              \def\theequation{\thesection.\arabic{equation}}}

\def\Label#1{\label{#1}%
                \smash{\hbox to0pt{\raise1ex\hbox{\tiny[#1]}\hss}}}
\def\noLabels{\let\Label=\label}

\def\tx{\tilde{x}}

\begin{document}

{\footnotesize
${}$
}

\bc

\vskip 10mm
\begin{center}\Large \bf
    The Vacuum Energy Problem in Quantum Gravity and the Masses of Elementary Particles
\end{center}
\vskip 10mm

\renewcommand{\thefootnote}{\fnsymbol{footnote}}

\centerline{{\bf
Djordje Minic${}^{}$\footnote{\tt dminic@vt.edu (Talk presented at BASIC 2023.) }
}}


{\it
${}$Department  of Physics, Virginia Tech, Blacksburg, VA 24061, U.S.A. \\
}

\ec


\begin{abstract}
This talk summarizes a new understanding of the cosmological constant problem, which  essentially
relies on a phase-space-like computation of the vacuum energy, both in the realm of quantum field theory coupled to gravity,
and in the realm of a consistent formulation of a quantum theory of gravity and matter, such as string theory,
combined with central properties of gravitational entropy as captured by the Bekenstein bound. 
This new understanding of the vacuum in a quantum theory of gravity and matter 
sheds light on the Higgs mass as well as the masses of quarks and leptons.

\end{abstract}


\renewcommand{\thefootnote}{\arabic{footnote}}


\section{Introduction}
The cosmological constant problem is considered to be one of the most important open problem in theoretical physics,
especially given the successes of Einstein's general theory of relativity
and the Standard Models of particle physics and of cosmology \cite{Weinberg:1988cp, 
Polchinski:2006gy}, as well as
the discovery of dark energy \cite{Riess:1998cb} 
which can be translated into a small and positive cosmological constant.
This result seems to be in blatant contradiction with canonical estimates of vacuum energy
in the context of effective field theory (EFT) \cite{Weinberg:1988cp,
Polchinski:2006gy}. 
We note that nature of the vacuum in gauge theories and gravity is much richer than previously thought, as illuminated recently in \cite{Strominger:2017zoo}, 
\cite{Freidel:2021wpl}. 
Also, the concepts such as gravitational entropy, holography and related quantum information theoretic ideas are important ingredients in our understanding of quantum theory of gravity \cite{Harlow:2022qsq}, and  
they complicate the application of EFT methods
\cite{Cohen:1998zx}.

The vacuum energy density problem is summarized in
\cite{Weinberg:1988cp, 
Polchinski:2006gy}.
In this talk we present a new understanding of the cosmological constant problem based on our recent work \cite{Freidel:2022ryr}.
We then apply the same logic to the Higgs mass as well as the masses of quarks and leptons.

\section{Vacuum Energy and Quantum Gravity}
Let us start  \cite{Polchinski:1998rq} 
with the QFT vacuum partition function of a free scalar field:
\begin{equation}
    Z_{vac} = \int D\phi e^{-\int \frac{1}{2}\phi (-\partial^2 + m^2) \phi} = \sqrt{\frac{\#}{ {\rm det}(-\partial^2 + m^2)}} = 
e^{-\frac{1}{2}{\rm Tr\, log}(-\partial^2 + m^2)}.
\end{equation}
Upon utilizing the fact 
that $-\frac{1}{2} \log(k^2+m^2) = \int \frac{{\rm d} \tau}{2\tau} e^{-(k^2+m^2) \tau/2} $, where the Schwinger parameter 
$\tau$ is a worldline parameter associated with a particle (quantum of the field $\phi$), and taking the trace, 
$\int \frac{{\rm d}^D k}{(2\pi)^D}\log(k^2 + m^2) = \int \frac{{\rm d}^{D-1}k}{(2\pi)^{D-1}} \frac{\omega_k}{2} $,
we are led to
the canonical expression for the vacuum energy which 
represents a sum (integral) over momentum space labels weighed by the canonical quantum expression for the zero point energy
$\rho_0 = \sum_k \frac{1}{2} \omega_k \to \rho_0 = \int \frac{d^3 k }{(2\pi)^3}\frac{1}{2} \omega_k,$
where $\omega_k = \sqrt{\vec{k}^2 +m^2}$ for a scalar field $\phi(x)$, and where $\hbar$ and $c$ are set to one. (The following discussion
could be generalized for other fields as well.)
This expression for the vacuum energy is naively quartically divergent in 4 spacetime dimensions and thus it has to be regulated.
It also states that
the vacuum energy density scales as the volume
of energy-momentum space $\Lambda_4$ (in what follows we consider the case of 4 spacetime dimensions, even though our discussion applies
more generally),
$\rho_0 \sim \Lambda_4 .$
When multiplied by the Newton gravitational
constant (as implied by Einstein's equations of general relativity that include the cosmological constant term), this regulated expression, usually cut-off by the Planck scale, gives a huge cosmological constant \cite{Weinberg:1988cp, 
Polchinski:2006gy}
$\lambda_{cc} \sim \rho_0 G_N \sim \rho_0 l_P^2,$
where $G_N \sim l_P^2$ is the four dimensional Newton's gravitational constant and $l_P$ is the corresponding Planck length.
This problem arises both in quantum field theory and in a quantum theory of gravity and matter, such as string theory  \cite{Polchinski:1998rq}. 
However, the observed cosmological constant is positive and small \cite{Riess:1998cb}.  
(Arguments like supersymmetry do not help with the offending scaling of the vacuum energy with the volume of energy-momentum space,
once supersymmetry is broken \cite{Polchinski:1998rq}. 
In the case of the unrealistic unbroken supersymmetry we get either the flat spacetime
with a zero cosmological constant, or an AdS space with a negative cosmological constant, in contradiction with observations \cite{Berglund:2022qsb}.)

By looking at the computation of the vacuum energy from the point of view of the evaluation of the vacuum partition function
$Z= \langle 0|\exp(-iHt) |0\rangle= \exp(-i \rho_0 V_4) $
(either in quantum field theory $Z= \exp(Z_{S^1})$, or in quantum gravity, such as string theory $Z= \exp(Z_{T^2})$), and by using the
fact that, for example \cite{Polchinski:1998rq}
\be
Z_{S^1} = V_4 \int \frac{d^4 k }{(2\pi)^4} \int \frac{d\tau}{2 \tau} \exp[-(k^2 +m^2)\tau/2] \equiv \int \frac{d\tau}{2 \tau} Z(\tau),
\ee
(and similarly for  $Z_{T^2}$, which, unlike the particle expression, is UV finite),
one obtains the following statement: The ratio of the vacuum partition function
for a particle (the excitation of a local quantum field $\phi(x)$) on a circle $Z_{S^1}$, or for a string on a torus $Z_{T^2}$, divided by the spacetime volume $V_4$
(for particles and field theory $\rho_0 \sim Z_{S^1}/V_4$, and for strings,  $\rho_0 \sim Z_{T^2}/V_4$)
leads always
to the scaling of the vacuum energy with the volume of energy-momentum space \cite{Polchinski:1998rq} 
$\rho_0 \sim \Lambda_4$. (In string theory, one has an infinite number of particles with masses $  m^2 = \frac{2}{\alpha'}(h+\bar h - 2)$,
where the conformal weights are constrained as $h = \bar h$. This constraint can be imposed via a delta function which 
eventually leads to a complex parameter $\tau$ on the torus $T^2$ \cite{Polchinski:1998rq}.)

The crucial observation of  \cite{Freidel:2022ryr} is as follows: the vacuum partition function for a particle on a circle, or a string on a torus, scales as the
(covariant) \emph{phase space volume} (the integrand of the partitition function being bound by one). Thus there is a bound for the
partition function given by the size of the phase space,
which, in the formulation based on a regulated phase-space of quantum theory (modular spacetime), is given by a number $N$  \cite{Freidel:2022ryr}. 
The number $N$ is the number of unit phase space boxes each of which has a characteristic spacetime scale $\lambda$ and the characteristic
energy-momentum space scale $\epsilon$ such that their product is the Planck constant  \cite{Freidel:2022ryr}
$\lambda \epsilon = 2 \pi.$
Here we note that modular spacetime is a notion of quantum spacetime, corresponding to a generic polarization of quantum theory  \cite{Freidel:2016pls}.
Note that in the context of modular spacetime, $x^\mu$ are the usual spacetime coordinates, while $p_\nu$ are dual coordinates conjugate to $x$. These coordinates are the eigenvalues of operators which do not commute, as implied by the canonical commutation relations
$[\hat{x}^\mu, \hat{p}_\nu]= i  \delta^\mu_\nu \hat I.$
The precise definition of modular spacetime presented in  \cite{Freidel:2016pls} is given as the space of all commuting subalgebras of the above Heisenberg algebra, which is equivalent to a discrete self-dual lattice in phase space. That is where the essential discreteness (modularity) of
phase space in the current computation arises.
The key idea of modular spacetime is the fact that it is possible to diagonalize
the operators $(\hat{x},\hat{p})$ as long as we restrict \cite{aharonov2008quantum,Freidel:2016pls} their eigenvalues to lie  inside a modular cell of size $\lambda$.  (The Heisenberg uncertainty principle  simply states that we cannot know  which specific modular cell inside phase space is chosen.)
A central feature of the modular cell can be rescaled in a way that respects the commutation relation through
$x\to a x,
p \to a^{-1} p$ .
In the limit of large expansion $a \to \infty$,  the dual dimension shrinks to zero and one recovers the usual spacetime.

In more detail, in 4 spacetime dimensions we have (for the simple case of $m=0$, but it can be demonstrated that the following is true more generally, even in the case of interacting quantum fields, and even at finite temperature, including cosmological phase transitions  \cite{Freidel:2022ryr}),
by discretizing phase space \cite{Freidel:2022ryr},
\be
Z(\tau) = \prod_{i=1}^4 \frac{1}{2\pi}\int_{-\infty}^\infty {\rm d}q_i \int_{-\infty}^\infty {\rm d}p_i e^{-\frac{p_i^2\tau}{2}} \to
Z(\tau) = \left(\frac{\lambda\epsilon}{2\pi} \sum_{k,\tilde{k}\in {\mathbf{Z}}} \int_0^1{\rm d} x{\rm d}\tilde{x} e^{-\frac{(k+\tilde{x})^2\epsilon^2\tau}{2}} \right)^4,
\ee
 where $p\to \epsilon \tilde{x}, q\to \lambda x$ 
with $\lambda\epsilon= 2 \pi$.
This procedure of {\it modular regularization} \cite{Freidel:2022ryr} results in
\be
Z(\tau) = \left( \sum_{k=0}^{N_q-1} \sum_{\tilde{k}=0}^{N_p-1} \int_0^1{\rm d} x{\rm d}\tilde{x} e^{-\frac{(k+\tilde{x})^2\epsilon^2\tau}{2}} \right)^4,
\ee
where $N_q,N_p$ count the number of unit cells in the spacetime and momentum space dimensions, respectively.
{\it Note that this expression is indeed bounded by the volume of phase space.}
Therefore, by using the basic fact that the product of the spacetime scale $\lambda$ and the energy-momentum scale $\epsilon$ is the Planck constant,
we get that the product of the (regularized) spacetime volume and the energy-momentum space volume (i.e. the phase space volume) is a number
$V_4 \Lambda_4 = N.$
Now we utilize the fact that if the number $N$ is equal to the number of spacetime boxes (by keeping the number of energy-momentum
space boxes equal to unity) then that number $N$ is the number of degrees of freedom in spacetime.
By using the Bekenstein bound \cite{Bekenstein:1980jp} (valid in any consistent quantum theory of gravity and matter) 
which tells us that the number of the degrees of freedom in spacetime is less
than the area of the boundary of spacetime (the cosmological horizon for de Sitter spacetime \cite{Berglund:2022qsb}) divided by the Newton gravitational
constant
$N \leq l^2/l_P^2,$
we deduce that the size of the volume of the energy-momentum space is 
$\Lambda_4 \leq \frac{1}{l^2 l_P^2},$
given that $V_4 \sim l^4$. Note that by construction $\Lambda_4 \sim \epsilon^4$, and $\lambda_{cc} \sim \epsilon^4 l_P^2$.
Thus the vacuum energy density, $\rho_0 \sim \Lambda_4$, is given by an expression that involves both the characteristic
IR ($l)$ and the UV ($l_P$) sizes, and the characteristic length scale $\lambda$ for the vacuum energy is given by the geometric
mean of the size of the Universe and the Planck scale
$\lambda = 1/{\epsilon} \sim \sqrt{l l_P}.$
This see-saw formula is consistent with observations if we take $l_P \sim 10^{-35}m$ and $l \sim 10^{27}m$,
which leads to the observed scale for the cosmological constant ($\lambda_{cc} \sim 1/l^2$), that is, the length scale 
$\lambda \sim 10^{-4}m$ or equivalently, 
the energy scale of $10^{-3}eV$.
This implies an extremely large number of phase-space (modular spacetime) boxes $N \sim l^2/l_P^2 \sim  10^{124}$.

Why would the number of modular cells (phase space boxes) $N$ have to be so large? $N$ has to be large so that the Universe is
stable: the fluctuation $\Delta$ for the size of an object built out of $N$ independent units 
scales as the inverse of the square root of the number of units, as is well known
from probability theory and statistical physics (this argument was used by Schr\"{o}dinger \cite{erwin} to explain
why atoms are so small compared to macroscopic bodies that have atomic structure),
$\Delta \sim \frac{1}{\sqrt{N}}$.
Also, the quartic root of that number $N$ is the number of ``atomic units'' in each of the spacetime directions ($i=0,1,2,3$),
$N_i \sim N^{1/4}$.
That number turns out to be $10^{31}$, which should be large because of stability ($\Delta_i \sim \frac{1}{\sqrt{N_i}}$), and is not that large when we remember the well-known Avogadro
number associated with the number of atoms in a characteristic macroscopic size of matter.

We emphasize the contextuality of the above argument. Note that $\epsilon$ is {\it not} a cut-off scale, as it would be
understood in EFT \cite{Cohen:1998zx}. 
Our argument regarding the scaling of the vacuum energy does not depend on
this scale, or the conjugate scale $\lambda$. Their product is always given by the Planck constant and they do not 
enter in the essential computation of the vacuum energy. If the context of the computation was changed (for example,
in the context of a cosmological evolution with phase transitions) the relevant scales would change contextually, even though the
essential computation would still have to be done in phase space (modular spacetime). Thus contextuality (a crucial property of quantum physics)
replaces arguments based on
the anthropic principle (i.e. statistical observer bias) \cite{Polchinski:2006gy}.

We also emphasize that EFT does not know above the number of phase space boxes $N$ and the Bekenstein bound.
Also, the vacuum partition function cancels in the computations of various correlation functions in EFT, and so EFT observables 
do not care about the vacuum partition function.
Also, the canonical quantum field theories (and string theory) live in classical spacetimes. One does computations in
spacetime or in its Fourier transform, the momentum space. One does not perform computations in phase space,
which is the key ingredient of our discussion.
(A historical remark: Veltman alluded to the need to use phase space in the elucidation of the cosmological constant problem
at the end of his Nobel Prize lecture in 1999 and in \cite{veltman}.)
We remark that by the above argument, the cosmological constant $\lambda_{cc} \sim 1/l^2$  is not dependent on any UV scale (and so, it is
radiatively stable) and for the large IR scale it goes to zero (thus, it is technically natural).

This discussion fits beautifully into our work \cite{Freidel:2021wpl} on metastrings and their zero modes metaparticles, propagating
in modular spacetime endowed with the Born geometry, a unification of the symplectic, doubly orthogonal and doubly-metrical geometry 
\cite{Freidel:2013zga}. 
(The Born geometry might be thought of as the relevant symmetry structure controlling
the above result for the cosmological constant.)
We have already alluded to modular spacetime as a generic polarization of quantum theory in the above discussion.
Another related concept is that of {metaparticles} introduced in \cite{Freidel:2018apz}, which generalizes the canonical concept of particles.
One way of thinking about metaparticles is that they can be obtained as the excitations of a modular spacetime.
On the other hand, metaparticles can be thought of as the metastring generalization of particle excitations  (for example,  particles
associated with the Standard Model (SM) fields)
\cite{Freidel:2015pka}. 
In both points of view, the fields that represent metaparticles are called {modular fields}, whose properties depend on a fundamental length scale $\lambda$ while at the same time preserving Lorentz covariance.
The modular fields which carry the metaparticle excitations are not simply functions of
spacetime coordinates $\phi(x)$ but functions $\phi(x, \tx)$ where $x^{\mu}$ are the usual spacetime coordinates while 
${\tx}_{\nu}$ are dual coordinates conjugate to $x$.
The fact that the modular fields and their metaparticle excitations manage to resolve this fundamental conundrum of quantum gravity \cite{Hossenfelder:2012jw} is one of the main reasons that we think they are of central importance in quantum gravity.
Metaparticles refer to the zero mode sector of metastring theory, a reformulation of string theory in which T-duality plays a central role \cite{Freidel:2015pka}.  
The metastring theory gives rise to a novel realization of quantum spacetime that is modular spacetime, and introduces new scales consistent with Lorentz covariance. In ordinary string theory, spacetime plays the role of an arena for particle dynamics, with the basic observables interpreted in terms of a particle S-matrix for the modes of the string  \cite{Polchinski:1998rq}.  
In metastring theory on the other hand, such a spacetime only appears through decoupling, in the large scale limit \cite{Freidel:2015pka}.  
Thus metastring theory is a quantum theory of gravity (or ``gravitization of quantum theory'' \cite{Freidel:2013zga}) that realizes a dynamical modular spacetime needed for the new understanding of the cosmological constant problem \cite{Freidel:2022ryr}.
As already stated, in the context of metastring theory apart from the usual spacetime coordinates $x^\mu$ one also encounters $\tilde{x}_\nu$, dual coordinates conjugate to $x$. These coordinates are the eigenvalues of operators which do not commute (in addition to the usual canonical commutators between coordinates and
dual momenta),
$[\hat{x}^\mu,\hat{\tx}_\nu]= i \pi \lambda^2  \delta^\mu_\nu \hat I.$
This commutation relation can be deduced through a careful analysis of the zero mode sector of string theory  
where $\lambda^2=\hbar\alpha'$ \cite{Freidel:2015pka}, but in general, the scale $\lambda$ depends on the context.
Once again, it is possible to diagonalize
the operators $(\hat{x},\hat{\tx})$ as long as we restrict \cite{aharonov2008quantum,Freidel:2016pls} their eigenvalues to lie  inside a modular cell of size $\lambda$.  
And once again, the modular cell can be rescaled in a way that respects the commutation relation through
$x\to b x,
\tx \to b^{-1} \tx$ .
Also, in the limit of large expansion $b \to \infty$,  the dual spacetime dimension shrinks to zero and one recovers the usual spacetime notions. This is the {decoupling} limit where a modular field $\phi(x, \tx)$ is restricted to the usual notion of a spacetime field $\phi(x)$ (for example, a SM field), as considered in the context of EFT. In this limit metaparticles (correlated objects of visible particles and their duals, which, in turn, represent natural candidates for dark matter \cite{Berglund:2022qsb})
are reduced to canonical particles.

\section{The Masses of Elementary Particles}

By repeating the above analysis of the vacuum energy in the context of metastring theory, one can also give a preliminary discussion on the gauge hierarchy problem
\cite{Berglund:2022qsb}, which in the realm of string theory (in contrast to EFT) is related to 
the cosmological constant problem \cite{Abel:2021tyt}.
The crucial input is provided by a careful computation (which can be realized in metastring theory) of a specific stringy partition function that obeys modular invariance \cite{Abel:2021tyt}, which results in a formula for the Higgs mass in string theory \cite{Abel:2021tyt}. Then by using intrinsic non-commutativity of the metastring, and by assuming that the cosmological (holographic/Bekenstein bound) scale
is replaced by the vacuum energy scale, we obtain the following seesaw formula for the Higgs ($m_H$) mass \cite{Berglund:2022qsb}
\be
m_{H} \sim \sqrt{m_{\Lambda} m_P} \sqrt{\langle X \rangle/8 \pi^2} \sim 125 GeV,
\ee
where $\langle X \rangle$ is an order one expression defined in \cite{Abel:2021tyt}, 
and where the cosmological constant seesaw formula derived in the previous section is
(given two scales, the Planck scale ($M_P$) $10^{19} GeV$ and the (Hubble) scale of the observed universe ($M$) 
$10^{-33}eV$) \cite{Freidel:2022ryr}
\be
m_{\Lambda} \sim \sqrt{M m_P} \sim 10^{-3}eV.
\ee

Note that the above resolution of the cosmological problem \cite{Freidel:2022ryr} implies that we know something important about the
vacuum of string theory, viewed as a quantum theory of gravity and matter. The computation of the Higgs mass reinforces that
intuition. But, if the vacuum is known, and one of the excitations (the Higgs) around the vacuum is known,
then the spectrum of perturbations of other excitations around the vacuum should be also known. In what follows we
want to shed light on the spectrum of other elementary particles in the Standard-Model-like matter sector of string theory
using the same logic that illuminated the cosmological constant problem and the Higgs mass. (The following is work in progress with
Per Berglund and Tristan H\"{u}bsch.)

First, as pointed out by Bjorken \cite{bjorken}, the masses of quarks and leptons could
be all parameterized in terms of a new scale $m \sim 10MeV$. This Bjorken-Zeldovich scale 
is given by the size of the universe and the Planck scale,
the two scales used in our argument for the cosmological constant, $l$ and $l_P$, according to
which $l_{cc}^2 \sim l l_P$
\be
l_{BZ}^3 \sim l l_P^2 \sim l_{cc}^2 l_P.
\ee
The most important point here is that given our $N \sim l^2/l_p^2$, the
Bjorken-Zeldovich scale $l_{BZ}^3 \sim l l_P^2 \sim l^3/N $ and
so 
$N \sim l^3/l_{BZ}^3$,
(as one would expect from an extensive non-gravitational entropy).
Therefore, given our $N$ and the Bekenstein bound for gravitational degrees of freedom,
and given the fact that in metastring theory the
matter and spacetime degrees of freedom are ``two sides of the same coin'' and 
given the extensive nature of entropy for the matter degrees of freedom
\be
N \sim l^3/l_{BZ}^3 \sim l^2/l_P^2,
\ee
we are able to {\it deduce} the Bjorken-Zeldovich scale,
$l_{BZ}^3 \sim l l_P^2 $ (corresponding to roughly $m \sim 10 MeV$, or equivalently, $10^{-14}m$).

Next, we use the masses of the heaviest quarks, the tau and the bottom, as well as the heaviest charged lepton, the tau, 
as the natural short distance scales. (For a computation of these masses in string theory, see \cite{Faraggi:1991be}.) These masses serve as analogs of the UV scale in our formula for the Higgs mass, whereas the
Bjorken-Zeldovich scale $m$ acts as the natural IR scale.
Note that the top mass $m_t$ is essentially tied to the Higgs
scale, which in turn is
given by the seesaw formula of the vacuum energy scale and the Planck scale.
(There is a  well-known argument based on criticality of the Standard Model that relates the masses of the top quark and the Higgs boson
\cite{Froggatt:1995rt}.)
Thus we obtain a seesaw relation for the charm mass
in terms of $m$ and $m_t$, that is analogous to the formula for the Higgs mass
\be
m_c \sim \sqrt{m m_t} \sim 1.1 GeV.
\ee
Next, by taking the bottom mass scale (for example, from the calculation of \cite{Faraggi:1991be}) 
as the analog of the top mass scale
and by using the same Bjorken-Zeldovich scale as the characteristic vacuum energy scale of matter,
we obtain the mass of the strange quark
\be
m_s \sim \sqrt{m m_b} \sim 170 MeV.
\ee
The up and down masses are given (according to Bjorken \cite{bjorken}) by essentially
the Bjorken-Zeldovich scale:
$m_u \sim m$ and  $m_d \sim m,$
but with some ``fudge factors'' so that $m_d > m_u$. The masses of the lightest quarks are
deduced from chiral perturbation theory
$m_u \sim 2 MeV, \quad m_d \sim 5 MeV.$
But according to our logic, apart from non-commutativity, we have T-duality, and
thus we need to use the second (T-duality like) seesaw relation.
This logic would lead to the following expressions
\be
m_u \sim m^2/m_c \sim m (m/m_c) \sim 10^{-2} m \sim 10^{-1} MeV.
\ee
Such a mass is a bit too small (off by a factor of 20), but it
is less (and that is important) than the mass of the down quark given as
\be
m_d \sim m^2/m_s \sim m (m/m_s) \sim 10^{-1} m 
\sim 1 MeV,
\ee
(also off by a factor of 5),
so that the proton is stable and the neutron decays.
To summarize: for the case of quark masses, given  
the heaviest, top and the bottom masses,  
the seesaw formulae of the first (non-commutative) and second (T-duality-like) kind produce the masses of the middle and the lightest quark generations.
One can also obtain the observed CKM matrix elements given the above formulas for the quark masses \cite{bjorken}.

Now let us examine the charged leptons: We need the analog of the top - and that is the tau.
From a naive stability analysis of the tau analog of the hydrogen atom,
the mass of the tau is expected to be of the order of 
the mass of the nucleus, i.e. a GeV.
Take the tau mass as given (again, from the calculation of \cite{Faraggi:1991be}), and repeat what we did with the heaviest quarks, using the tau mass and the Bjorken-Zeldovich scale to deduce the mass of the muon
\be
m_{\mu} \sim \sqrt{m m_{\tau}} \sim 110MeV.
\ee
Using the second seesaw (of the T-duality kind) as before, 
we also obtain the electron mass, given the calculated muon mass
\be
m_e \sim \frac{m^2}{m_{\mu}} \sim 450KeV.
\ee

So, there are 3 generations of leptons and quarks, according to this proposal,
because of the set up offered
by the dual space, the modular spacetime Born geometry, and ultimately the metastring,
or essentially, the intrinsic non-commutativity and covariant T-duality of the metastring.
In other words, apart from the usual classical
spacetime, there is a dual spacetime (coming from
the double metric of Born geometry) and the
T-dual of that spacetime (coming from the 
bi-orthogonal metric). That is why the
SM modular field theory (embedded in metastring theory) naturally reproduces
the 3 generation SM in classical spacetime.
The masses of the other two generations
(starting from the heaviest quarks and leptons)
are fixed by non-commutativity and T-duality,
in analogy with the reasoning that gives the Higgs mass and the cosmological constant.
All of these formulas are seesaw-like and contextual.
All of them ultimately reduce to the
IR size of the universe and the UV Planck length.

Finally, we consider neutrino masses. 
Given the dimension 5 (Weinberg) operator in the Standard Model \cite{Weinberg:1979sa}
(implying Majorana masses as well),
we set the heaviest (''tau'') neutrino mass to be
\be
m_3 \sim m_H^2/m_{SM} \sim 10^{-1} eV,
\ee
where the SM scale is given by a ``would-be-unification scale'' of
the SM couplings - $10^{16} GeV$, and $m_H$ is the Higgs scale of
around $1 TeV$. 
The middle (''muon'') neutrino mass is given
by a seesaw formula involving a low vacuum energy scale.
However, the neutrinos (unlike all quarks and charged leptons) do not get
their masses from the Higgs mechanism, and thus the vacuum scale
cannot be the Bjorken/Zeldovich scale (used
for the quarks and charged leptons) but the only other vacuum scale,
that is the cosmological
vacuum scale associated with the cosmological constant, which we already know. Thus
\be
m_2 \sim \sqrt{m_{\Lambda} m_3} \sim 10^{-2} eV.
\ee
(Note that it has been argued in \cite{Aydemir:2017hyf}
that this mass value
is also natural if one examines a dimension 6 operator, an analog of Weinberg's dimension 5 operator, in which a neutrino 
could obtain its mass from a
fermionic condensate controlled by the
Bjorken-Zeldovich scale, with a
cutoff provided by the electroweak scale,
so that $m_2 \sim m_{BZ}^3/m_{H}^2$.)
Finally, the lightest (``electron'') neutrino mass is
given by the T-duality seesaw formula
\be
m_1 \sim m_{\Lambda}^2/m_2 \sim 10^{-4} eV.
\ee
According to the Particle Data Book (https://pdg.lbl.gov/) the sum of neutrino masses (coming from cosmology) is bounded by
$10^{-1} eV$ which is satisfied by the above normal hierarchy of neutrino masses. 
Also, these values satisfy the constraint on the square of the differences of masses ($10^{-2} eV^2 - 10^{-5} eV^2$) coming from neutrino oscillation experiments.

Note that all these masses of quarks and leptons (and the value of the cosmological constant and the Higgs mass) are upper bounds (the bound for the up and down quark masses being given by the Bjorken-Zeldovich scale) and thus we expect an
attractor mechanism (as in \cite{gcode}) that would ''glue'' all these 
values to their upper bounds. This would be consistent with the existence of a self-dual fixed point in metastring theory \cite{Freidel:2015pka} that could explain
the apparent criticality of the Standard Model parameters \cite{Froggatt:1995rt}.

\section{Conclusion}
In conclusion, in this talk the recent understanding of the cosmological constant problem \cite{Freidel:2022ryr} was presented
along the extension of the same argument to the problem of the Higgs mass \cite{Berglund:2022qsb} and the masses of quarks and leptons.
We remark that our result relating the number of phase space boxes to the Bekenstein bound ($N \sim l^2/l_p^2$), can
be used for black holes (with $l$ being the size of the black hole horizon $l_{bh}$) where it is naturally related to \cite{Bekenstein:1995ju}.
In our case the relevant quantization number $N_{bh} \sim l_{bh}^2/l_P^2$ for black holes is of the order of $10^{80}$, and a possible observable feature of this quantization $l_{bh}^2 \sim N l_P^2$ might be via the ``gravitational wave echoes'' \cite{Cardoso:2019apo} (in the ``quantum chaos'' phase, given
the enormous value of $N$).
This would be a fascinating empirical signature of one of the key elements in our discussion. Finally we note that, apart from metaparticles, the
most spectacular empirical prediction of our approach to quantum gravity as ``gravitization of the quantum'' 
\cite{Freidel:2013zga} is represented by the phenomena of intrinsic triple and higher order interference \cite{Berglund:2023vrm}.

\noindent
{\bf Acknowledgments:}
Many thanks to L, Freidel, J. Kowalski-Glikman and R. G. Leigh as well as P. Berglund, A. Geraci, T. H\"{u}bsch and D. Mattingly  
for insightful collaborations and T. Curtright and E. Guendelman for the invitation to BASIC 2023.
This work is supported by the  Julian  Schwinger Foundation and
the U.S. Department of Energy (under contract DE-SC0020262).


\begin{thebibliography}{100}


\bibitem{Weinberg:1988cp}
S.~Weinberg,
Rev. Mod. Phys. \textbf{61} (1989), 1-23;
S.~Weinberg,
[arXiv:astro-ph/0005265 [astro-ph]];
E.~Witten,
[arXiv:hep-ph/0002297 [hep-ph]];
T.~Padmanabhan,
Phys. Rept. \textbf{380} (2003), 235-320;
A.~Padilla,
[arXiv:1502.05296 [hep-th]].

%
\bibitem{Polchinski:2006gy}
J.~Polchinski,
[arXiv:hep-th/0603249 [hep-th]].



\bibitem{Riess:1998cb}
A.~G.~Riess \textit{et al.},
Astron. J. \textbf{116}, 1009-1038 (1998);
S.~Perlmutter \textit{et al.},
Astrophys. J. \textbf{517}, 565-586 (1999).


\bibitem{Strominger:2017zoo}
A.~Strominger,
[arXiv:1703.05448 [hep-th]];
%
W.~Donnelly and L.~Freidel,
JHEP \textbf{09}, 102 (2016); 
L. Ciambelli, R.G. Leigh and P.-C. Pai, PRL 128 (2022) 171302;
L. Freidel,  [arXiv:hep-th/2111.14747].
%

\bibitem{Freidel:2021wpl}
L.~Freidel, J.~Kowalski-Glikman, R.G.~Leigh and D.~Minic,
Int. J. Mod. Phys. D \textbf{30}, no.14, 2141002 (2021);
%
P.~Berglund, L.~Freidel, T.~Hubsch, J.~Kowalski-Glikman, R.~G.~Leigh, D.~Mattingly and D.~Minic,
[arXiv:2202.06890 [hep-th]].



\bibitem{Harlow:2022qsq}
D.~Harlow, 
\textit{et al.}
[arXiv:2210.01737 [hep-th]].


%
\bibitem{Cohen:1998zx}
A.~G.~Cohen, D.~B.~Kaplan and A.~E.~Nelson,
Phys. Rev. Lett. \textbf{82}, 4971-4974 (1999). See also,
%
J.~F.~Donoghue,
Phys. Rev. D \textbf{104}, no.4, 045005 (2021).


\bibitem{Freidel:2022ryr}
L.~Freidel, J.~Kowalski-Glikman, R.~G.~Leigh and D.~Minic,
[arXiv:2212.00901 [hep-th]], and [arXiv:2303.17495 [hep-th]].
Also, work in progress.


%
\bibitem{Polchinski:1998rq}
J.~Polchinski,
{\it String theory. Vol. 1: An introduction to the bosonic string},  Cambridge University Press, 1998.
%
J.~Polchinski,
Commun. Math. Phys. \textbf{104}, 37 (1986).


%
\bibitem{Berglund:2022qsb}
P.~Berglund, T.~H\"ubsch and D.~Minic,
[arXiv:2212.06086 [hep-th]]. See also,
[arXiv:2111.14205 [hep-th]]; LHEP \textbf{2021}, 186 (2021); Int. J. Mod. Phys. D \textbf{28}, no.14, 1944018 (2019);
Phys. Lett. B \textbf{798}, 134950 (2019).




\bibitem{Freidel:2016pls}
  L.~Freidel, R.~G.~Leigh and D.~Minic,
  Phys.\ Rev.\ D {\bf 94}, no. 10, 104052 (2016).

\bibitem{aharonov2008quantum}
Y.~Aharonov and D.~Rohrlich, {\em Quantum Paradoxes: Quantum Theory for the
  Perplexed}.
\newblock John Wiley \& Sons, 2008.



\bibitem{Bekenstein:1980jp}
J.~D.~Bekenstein,
Phys. Rev. D \textbf{23}, 287 (1981).


\bibitem{erwin} 
E. Schr\"{o}dinger, {\it What is life?}, Cambridge University Press 1992.

\bibitem{veltman}
M. Veltman, Acta. Phys.  Pol. B25 (1994) 1399.


\bibitem{Freidel:2013zga}
  L.~Freidel, R.~G.~Leigh and D.~Minic,
  Phys.\ Lett.\ B {\bf 730}, 302 (2014);
  Int.\ J.\ Mod.\ Phys.\ D {\bf 23}, no. 12, 1442006 (2014).




\bibitem{Freidel:2018apz}
  L.~Freidel, J.~Kowalski-Glikman, R.~G.~Leigh and D.~Minic,
  Phys.\ Rev.\ D {\bf 99}, no. 6, 066011 (2019).



\bibitem{Freidel:2015pka}
  L.~Freidel, R.~G.~Leigh and D.~Minic,
  JHEP {\bf 1506} (2015) 006;
  J.\ Phys.\ Conf.\ Ser.\  {\bf 804}, no. 1, 012032 (2017).
  JHEP {\bf 1709}, 060 (2017);
  Phys.\ Rev.\ D {\bf 96}, no. 6, 066003 (2017).




\bibitem{Hossenfelder:2012jw}
S.~Hossenfelder,
Living Rev. Rel. \textbf{16} (2013), 2;
A.~Addazi, \textit{et al.}
Prog. Part. Nucl. Phys. \textbf{125}, 103948 (2022).




 



%
\bibitem{Abel:2021tyt}
S.~Abel and K.~R.~Dienes,
Phys. Rev. D \textbf{104}, no.12, 126032 (2021).


\bibitem{bjorken}
J. D. Bjorken, Annalen der of Physik 525, A67, (2013). 

\bibitem{Faraggi:1991be}
A.~E.~Faraggi,
Phys. Lett. B \textbf{274}, 47-52 (1992); 
Nucl. Phys. B \textbf{487}, 55-92 (1997).

%
\bibitem{Froggatt:1995rt}
C.~D.~Froggatt and H.~B.~Nielsen,
Phys. Lett. B \textbf{368}, 96-102 (1996).
See also, J.~F.~Donoghue, K.~Dutta and A.~Ross,
Phys. Rev. D \textbf{73}, 113002 (2006); J.~Khoury and S.~S.~C.~Wong,
[arXiv:2205.11524 [hep-th]].


%
\bibitem{Weinberg:1979sa}
S.~Weinberg,
Phys. Rev. Lett. \textbf{43}, 1566-1570 (1979).

%
\bibitem{Aydemir:2017hyf}
U.~Aydemir,
Universe \textbf{4}, no.7, 80 (2018)
[arXiv:1704.06663 [hep-ph]].

\bibitem{gcode}
J-A. Argyriadis, Y-H. He, V. Jejjala and D. Minic,
Phys. Rev. E 103, 052409 (2021).


\bibitem{Bekenstein:1995ju}
J.~D.~Bekenstein and V.~F.~Mukhanov,
Phys. Lett. B \textbf{360}, 7-12 (1995).

%
\bibitem{Cardoso:2019apo}
V.~Cardoso, V.~F.~Foit and M.~Kleban,
JCAP \textbf{08}, 006 (2019), and references therein.

%
\bibitem{Berglund:2023vrm}
P.~Berglund, A.~Geraci, T.~H\"ubsch, D.~Mattingly and D.~Minic,
[arXiv:2303.15645 [gr-qc]], and 
P.~Berglund, T.~H\"ubsch, D.~Mattingly and D.~Minic,
Int. J. Mod. Phys. D \textbf{31}, no.14, 2242024 (2022)
[arXiv:2203.17137 [gr-qc]].



\end{thebibliography}
\end{document}